\documentclass[%
  reprint, prl,
superscriptaddress,
 amsmath,amssymb,
 aps,
]{revtex4-2}

\usepackage[
bookmarks=true,
colorlinks,
linkcolor=blue,
urlcolor=blue,
citecolor=olive,
plainpages=false,
pdfpagelabels,
final,
breaklinks=true
]{hyperref}

\usepackage{graphicx}
\usepackage{dcolumn}
\usepackage{bm}
\usepackage{soul}

\usepackage{xcolor}

\begin{document}

\preprint{APS/123-QED}

\title{Field-driven helicity in solid-state high-harmonic generation}

\author{Carlos Batista}
\affiliation{Departamento de F\'isica, \'Area de F\'isica, Universidad de Panam\'a,
Ciudad Universitaria 3366 Octavio Mendez Pereira, Panama}
\author{Jean Paul Menotti}
\affiliation{Departamento de F\'isica, \'Area de F\'isica, Universidad de Panam\'a,
Ciudad Universitaria 3366 Octavio Mendez Pereira, Panama}

\author{Dasol Kim}
\affiliation{Department of Materials Science and Engineering, Pohang University of Science and Technology, Pohang 37673, Korea}

\author{Bikash Kumar Das}
\affiliation{Theoretisch-Physikalisches Institut, Friedrich-Schiller-Universität Jena, Max-Wien-Platz 1, D-07743 Jena, Germany}

\author{Wenlong Gao}
\affiliation{$^7$Eastern Institute of Technology, Ningbo, China}

\author{Alexis Chac\'on}
\affiliation{Departamento de F\'isica, \'Area de F\'isica, Universidad de Panam\'a,
Ciudad Universitaria 3366 Octavio Mendez Pereira, Panama}
\affiliation{Sistema Nacional de Investigación de Panamá,  Building 205, Ciudad del Saber, Clayton Panamá}
\affiliation{Centro de Investigaci\'on con T\'ecnicas Nucleares, Universidad de Panam\'a, Panama}
\affiliation{Parque Cient\'ifico y Tecnol\'ogico, Universidad Aut\'onoma de Chiriqu\'i, Ciudad Universitaria, David, Panama}
\email{alexis.chacon-s@up.ac.pa}

\author{Camilo Granados}
\affiliation{Eastern Institute of Technology, Ningbo, China}
\email{cagrabu@eitech.edu.cn}

\date{\today}

\begin{abstract}
The polarization state of light plays a central role in strong-field light--matter interactions and is widely used to probe electronic structure in solids via high-order harmonic generation (HHG). In particular, helicity-resolved HHG has been interpreted as a fingerprint of crystal symmetry and topology. Here, we demonstrate deterministic and continuous control of harmonic helicity in solids using polarization-crafted beams, formed by two orthogonally polarized pulses with a controlled time delay. By tuning this delay, the polarization state of individual harmonics can be driven from linear to circular, independent of the material under investigation. We show that this behavior is robust across systems with distinct symmetry and topology, and originates from the sub-cycle modulation of the light--matter interaction mediated by the dipole coupling. Furthermore, the orthogonal configuration allows to break the dynamical symmetry of the light-matter interaction which is manifested in the generation of otherwise forbidden harmonics  under standard selection rules.. These results establish harmonic helicity as a field-controlled observable rather than a direct material fingerprint. 
\end{abstract}

\maketitle

\emph{Introduction--} 
The polarization state, in particular the helicity, of light is a fundamental degree of freedom in strong-field light-matter interactions and provides direct access to ultrafast electronic dynamics in solids. In high-order harmonic generation (HHG), polarization-resolved measurements have emerged as a powerful tool to probe crystal symmetries, Berry curvature effects, and topological properties, while enabling the generation of circularly polarized and attosecond light sources \cite{HHGAniso,TopHHG2,ChristianNatPhoto2021,ChaconPRB2020,SL3,Pol2,Pol3,Pol4,PolSolids,Langer2017}.

Helicity-resolved HHG has been widely interpreted as a fingerprint of material properties. In systems with broken inversion or time-reversal symmetry, circularly polarized driving fields can induce asymmetric harmonic emission, linking helicity to the underlying electronic structure \cite{TopoHHG,3DTopInsu,TopHHG2}. This connection has motivated the use of harmonic helicity and circular dichroism as probes of topological phases and chiral responses in solids \cite{ChiralSolidsHHG}. More broadly, helicity-dependent effects have also been explored using structured and helical light fields in atoms, molecules, and amorphous systems \cite{ForbesPRA2019,ForbesOptLett2018,JainNatComm2024,JainNatComm2024,Helical_dichro2,Helical_dichro3}, highlighting helicity as a versatile observable beyond crystalline order.

However, recent studies have raised fundamental questions about the universality of such interpretations \cite{NeufeldPRX2023}. In particular, the extent to which the helicity of emitted harmonics is determined by the material, as opposed to the driving field, remains unclear. Disentangling these contributions is essential for establishing helicity-resolved HHG as a reliable probe of electronic structure.

Here, we demonstrate deterministic and continuous control of the helicity of high-order harmonics in solids using polarization-crafted beams (PCBs), formed by two orthogonally polarized pulses with a controlled time delay \cite{PCB1,GranadosEPJD2024}. By tuning this delay, the polarization state of the emitted harmonics can be continuously altered from linear to circular, independent of the material under investigation. We show that this behavior is robust across systems with distinct symmetry and topology, and originates from the sub-cycle modulation of the light-matter interaction mediated by the dipole coupling. Our results establish harmonic helicity as a field-controlled observable rather than only a direct material fingerprint. Furthermore it provides a route to polarization-tailored attosecond sources in solids.

\begin{figure}[htbp]
    \centering
    \includegraphics[width=0.5\textwidth,trim=6cm 2cm 5cm 2cm,clip]{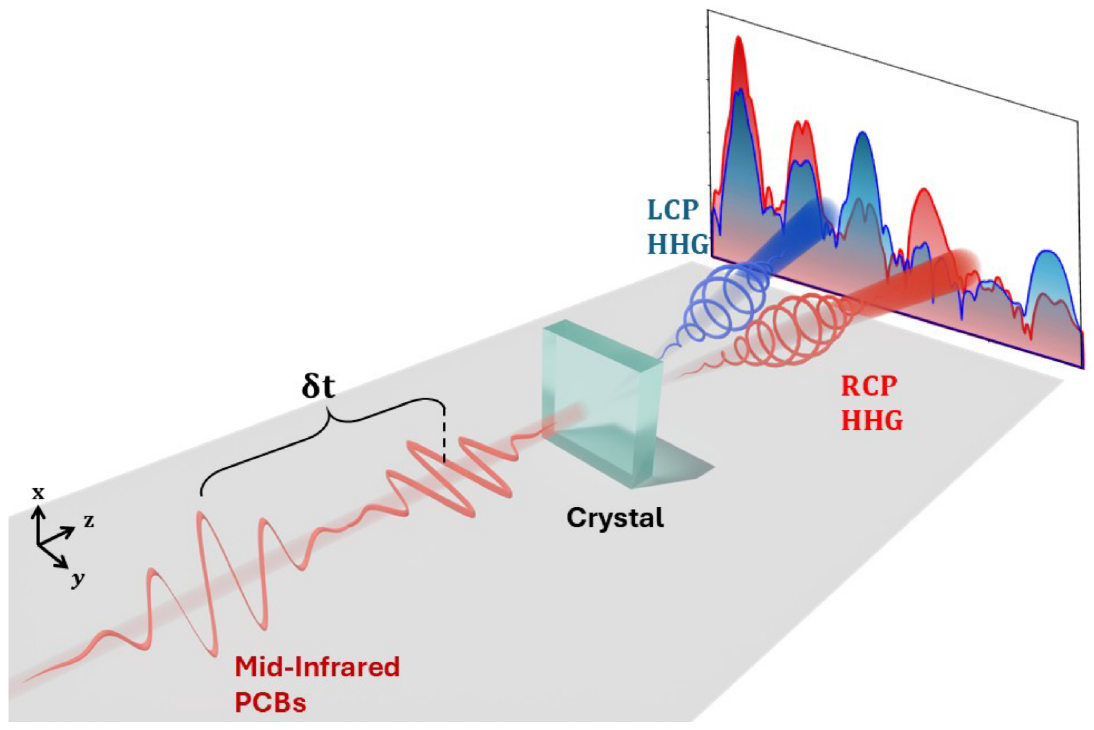}
    \caption{
    Schematic representation of the high-harmonic generation in solids driven by PCBs. In this configuration, two strong, ultrashort mid-infrared (MIR) pulses with orthogonal polarizations and a controlled time delay $\delta t$ are focused on the crystal. This allows us to control the harmonics helicity via the inter and intraband currents.} 
    \label{fig:intraband_interband_helicity}
\end{figure}

In Fig.~\ref{fig:intraband_interband_helicity}, we present a schematic representation of the HHG process driven by the PCBs. In short, the HHG spectra are simulated for each time delay between the pulses, $\delta t$, which is continuously tune. This time delay results is harmonics with controllable helicity. Here, right- and left-circular (helicity) components are define as $J_{RCP/LCP}=(J_x\pm i J_y)/\sqrt{2}$, from where we calculate the helicity-resolved spectra.

\emph{Polarization crafted beams--} We start by revisiting the PCBs  which are produced by the superposition of two electromagnetic (EM) fields sharing the same frequency and with perpendicular linear polarization relative to each other. Mathematically, the field distribution of the PCBs is given by:
\begin{equation} \label{MathPCBs}
\vec{E}_t(t,\delta t,\phi_x,\phi_y)=E_x(t,\phi_x)\hat{x}+E_y(t-\delta t,\phi_y)\hat{y}.
\end{equation}
\noindent Here, the time delay between the pulses is represented by $\delta t$ and the carrier-envelope phase of the individual pulses by $\phi_i$.~Furthermore, the components of the total EM field are ${E}_i(t,\phi_i)$, where $\hat{i}$ represents the components in $\hat{x}\,,\hat{y}$. Each component corresponds to fundamental fields with a Gaussian envelope and equal field amplitude, $E_0$. The linear- and circular-like polarization states of the superposed beams can be controlled by changing either $\delta t$ or $\phi_i$. The time delay between the two components defines the polarization of the driving field, i.e., for $\delta t=\pm T_0/2$ ($T_0$ $\equiv$ laser period) the driver polarization is a linear-like, while for  $\delta t= \pm T_0/4$, we obtain a circular-like polarization pulse. We note that intermediate states of polarization are also possible for different time delay values.~Additionally, the sign of the time delay, $\delta t$, controls the handedness of the pulses, i.e., $E_x(t,\phi_x)\hat{x}+E_y(t-\delta t,\phi_y)\hat{y}$ results in a right-circularly polarized (RCP) pulse and $E_x(t,\phi_x)\hat{x}+E_y(t+\delta t,\phi_y)\hat{y}$ in a left-circularly polarized (LCP) pulse. 

\emph{Harmonic generation in Graphene and Bi$_2$Se$_3$ --}
We first consider high-order harmonic generation driven by conventional circularly and linearly polarized fields in graphene and Bi$_2$Se$_3$. These calculations are based on the time-dependent density matrix formalism \cite{ChaconPRB2020} and also serve as a benchmark for our theoretical framework (see Supplementary material). These calculations reproduce the ellipticity-dependent harmonic yield and its connection to crystal rotational symmetry, in agreement with previous theoretical and experimental results \cite{GraEllip}. 

\begin{figure}[h!]
\centering 
\includegraphics[width=1\linewidth]{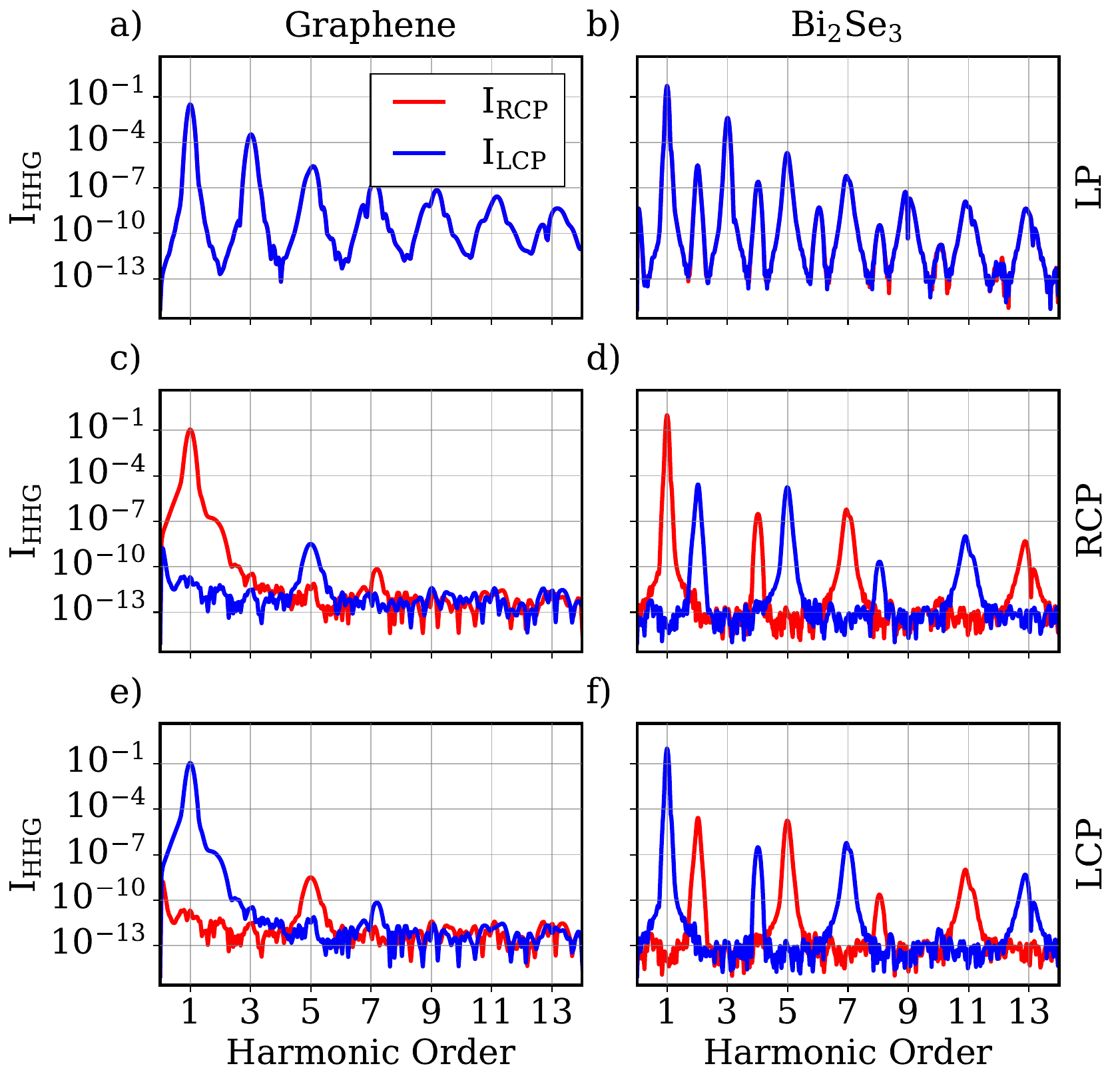}
\caption{High-order harmonic generation from conventional fields.~The spectra for graphene, is shown in panels a), c), and e), while the spectra for  Bi$_2$Se$_3$ is shown in panels b), d) and f). The different spectra results from a driver with linear-polarization (LP), RCP, and LCP (from top to bottom).~ Red and blue lines correspond to the RCP and LCP decompositions of emitted harmonics. For graphene, we used the Haldane-like model with nearest-neighbor hopping $t_1 = 0.1029~\text{a.u. } (\approx 2.80~\text{eV})$, next-nearest-neighbor hopping $t_2 = 0~\text{a.u.}$, on-site potential ratio $M_0/t_2 = 0$, and magnetic flux $\phi_0 = 0$. For Bi$_2$Se$_3$, we focus exclusively on the topological surface states, with intralayer hopping parameters $B_0 = 0.0164~\text{eV}$ and $B_{11} = 0.1203~\text{eV}$. The laser parameters are presented in the text.}
\label{Fig1}
\end{figure}

Figure~\ref{Fig1} summarizes the results. Linearly polarized driving fields generate spectra shown in panels (a) and (b), while right- and left-circularly polarized drivers produce the spectra in panels (c)--(f). In graphene, the preservation of the inversion symmetry enforces odd harmonics only, whereas in Bi$_2$Se$_3$ surface states allow both even and odd harmonics to appear due to broken inversion symmetry.  These features are captured by the selection rules:
\begin{eqnarray}
    q = nj + \sigma_{\textrm{HHG}},
\end{eqnarray}
where $q$ is the harmonic order, $j \in \mathbb{N}$, and $\sigma_{\textrm{HHG}}=\{-1,+1\}$. For a system with $C_N$ rotational symmetry, this reduces to the well-known constraint that only $nj \pm 1$ harmonics are allowed, consistent with the dynamical symmetry $\hat{U}=\hat{R}_{2\pi/n}\hat{T}_{T/n}$ satisfying $\hat{U}\hat{H}(t)\hat{U}^\dagger=\hat{H}(t)$ \cite{Cohen_Symmetries,SL5}.

In addition, the harmonic polarization follows parity-imposed helicity rules, where successive allowed harmonics exhibit alternating helicity depending on the driving-field handedness. Importantly, while the helicity distribution changes between right- and left-circular driving fields, the total harmonic yield remains unchanged, resulting in vanishing net helicity for conventional circular driving fields. This reflects the fact that symmetry constraints determine the allowed emission channels but do not provide independent control over their helicity population.

This limitation motivates the use of structured driving fields. In the following, we introduce polarization-crafted beams (PCBs) \cite{PCB1,GranadosEPJD2024} driven solid-state HHG, which provide a direct manipulation of the harmonic helicity beyond symmetry-imposed constraints.

\emph{Helicity control--} The PCBs allow for the control over the electron dynamics in the solid which is mapped by the helicity of the emitted harmonics and it is independent of the target material. To demonstrate this, we simulated the HHG process in a large variety of solids, including trivial and topological materials. For the numerical simulations, we used PCBs centered at $7.5~\mu$m, and with a field amplitude of $E_0 = 0.00054~\text{a.u.}$ ($I_0 = 1.0 \times 10^{10}~\text{W/cm}^2$), and a pulse duration of 8 optical cycles. The time delay values were set to $\delta t = \{ T_0/4, T_0/2\}$. 

\begin{figure}
\centering 
\includegraphics[width=1\columnwidth,trim=6cm 0cm 5.5cm 0cm,clip]{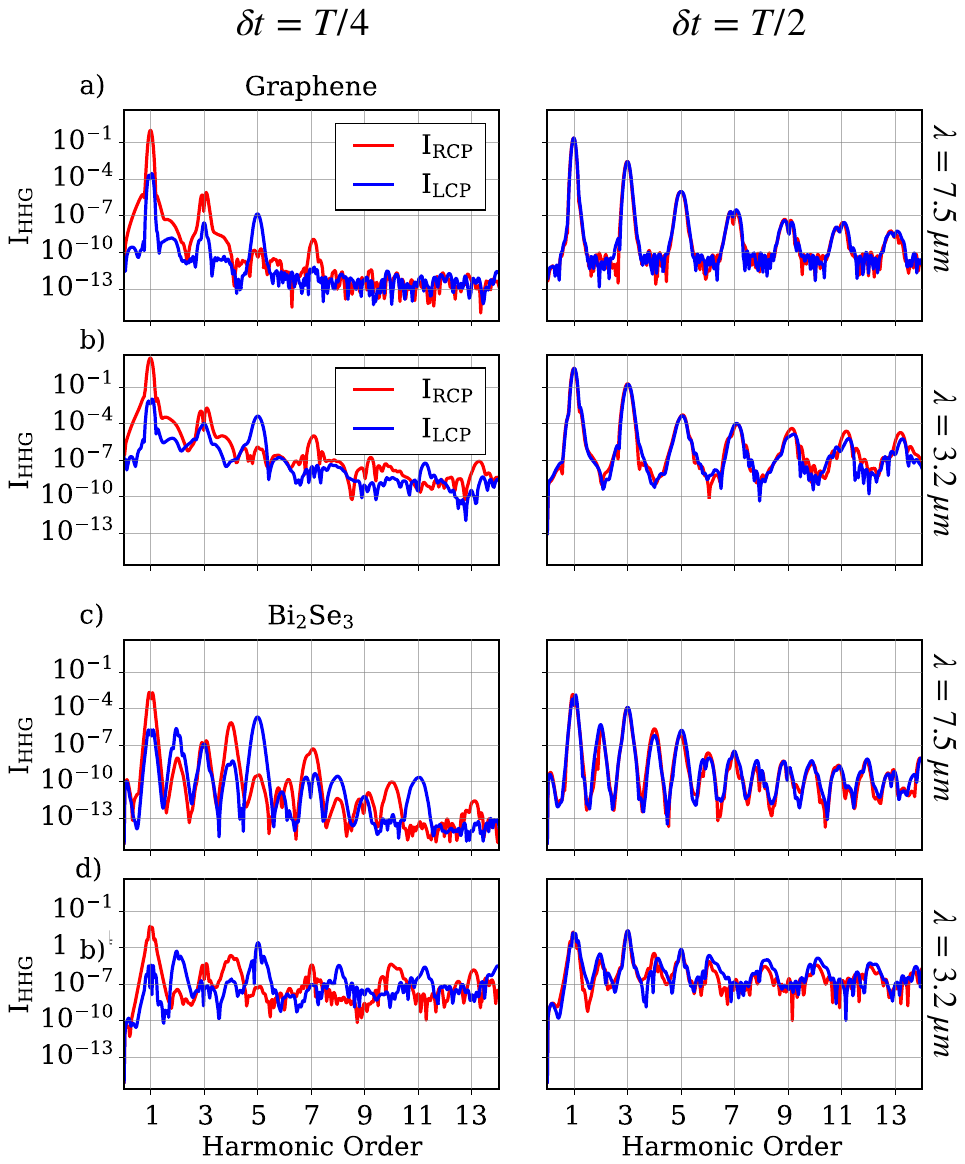}
\caption{High-harmonic spectra for different time delays. Panels a) and b) we show the HHG spectra for graphene and panel c) and d) for Bi$_2$Se$_3$ surface states. Furthermore, we calculated the spectra for two different wavelengths, $\lambda = 7.5~\mu$m and $\lambda = 3.2~\mu$m .~We observe a clear difference between the RCP and LCP harmonic yields, giving rise to a nonzero helicity for the time delay $\delta t = T_0/4$. However, for $\delta t = T_0/2$ there are no significant differences, consistent with the case of a linearly polarized driver. For both materials, we use the same parameters as in Fig.~\ref{Fig1}. See the text for more details. } 
\label{Fig2}
\end{figure}

\noindent An example of the resulting harmonic spectra for the different materials is shown in Fig.~\ref{Fig2}, where the spectra presented in the left panel were calculated with a time delay $\delta t = T_0/4$ while the spectra in the right panel with $\delta t = T_0/2$.~In each panel, the color code indicates RCP in red and LCP in blue. Importantly, circular-like polarization leads to asymmetric RCP and LCP components, which are time-delay dependent (see below).~We used the following definition for the right and left polarized PCBs: for $\delta t>0$ the pulses are LCP, while $\delta t <0$ corresponds to RCP and with the convention of Eq.~\ref{MathPCBs}. 

\begin{figure}
\centering 
\includegraphics[width=1\linewidth, trim= 0cm 0cm 0cm 0cm, clip]{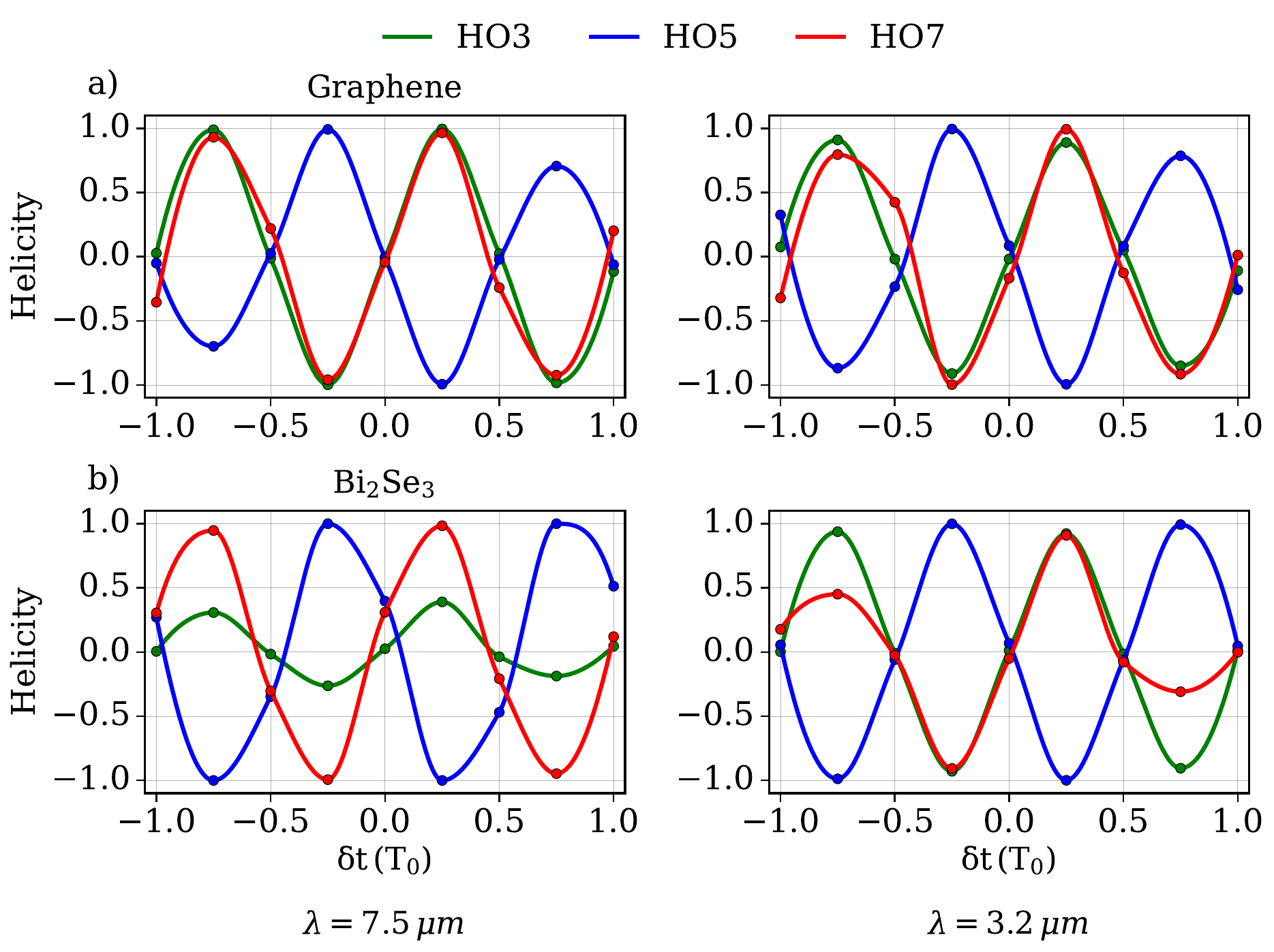}
\caption{
Time-delay dependent helicity. In Panels (a) and (b), we show the resulting time-delay dependent helicity for graphene and Bi$_2$Se$_3$ and for two different wavelengths, $\lambda = 7.5~\mu$m and $\lambda = 3.2~\mu$m(see Supplementary material for the other solid-state materials). Here, we demonstrate the control over the helicity as a function of the time delay between the pulses, $\delta t$, for different materials. Notice that the solid lines are used here only for guiding the eye, while the points indicate the exact temporal delay, $\delta t$, used to calculate the harmonic spectra.}
\label{TDHelicity}
\end{figure}

From the harmonic spectra we can extract the helicity for each time delay. The results are presented in Fig.~\ref{TDHelicity} and for the harmonic orders 3, 5 and 7. The changes in the helicity demonstrate two different effects: (i) the polarization state for the harmonics emitted in the different materials is controllable by the time delay, and (ii) the results are similar for both the materials. Similar results were obtained for other materials, including the Kane-Mele model which represents a Topological material, as presented in the Supplementary material. 

\emph{Discussion--} The time-delay dependent helicity, demonstrated for all the harmonics in Fig.~\ref{Fig2}, appears to be a universal feature of the solid-state HHG.~The simple arrangement provided by the PCBs is a powerful tool to control light's helicity as well as for controlling the inter and intraband electronic dynamics. Importantly, polarization-crafted beams induce a controlled breaking of the dynamical symmetry $\hat{U}=\hat{R}_{2\pi/n}\hat{T}_{T/n}$, which manifests as a violation of the standard selection rules $q = nj \pm 1$. This symmetry breaking is directly observed through the appearance and enhancement of harmonics that are suppressed under ideal circular driving, such as the third harmonic in both graphene and Bi$_2$Se$_3$.

\begin{figure}[htbp]  
    \centering
    \includegraphics[width=\columnwidth]{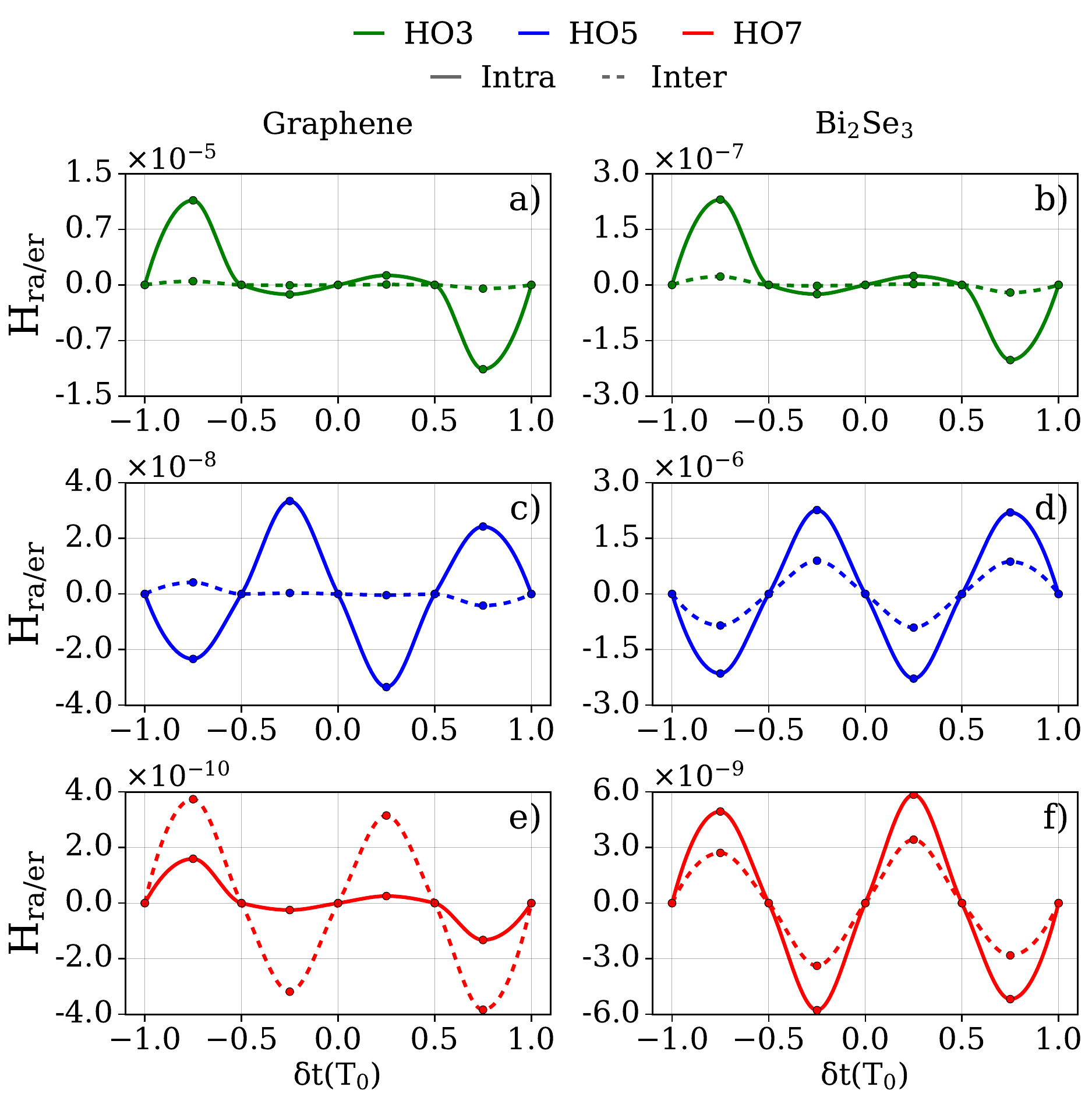} 
    \caption{
Intraband (ra) and interband (er) pump-probe time-delay-dependent unnormalized helicity $H_{\rm ra/er}$. The left column (panels a), c), and e)) corresponds to graphene, while the right column (panels b), d), and f)) corresponds to Bi$_2$Se$_3$. Rows display HO3 (a), b)), HO5 (c), d)), and HO7 (e), f)). The solid lines correspond to the intraband helicity, while dashed lines correspond to the interband helicity. The results demonstrate that the helicity control is associated with the oscillations in the inter and intraband currents induced by the PCBs. Furthermore, the differences between the intraband values for left ($\delta t < 0$) and right polarized PCBs ($\delta t > 0$) explains the origin of time-delay dependent helicity.}
    \label{Disc1}
\end{figure}

To further understand the microscopic origin of the helicity modulation, we define a helicity-like quantity for the individual inter and intraband contributions to the total harmonic spectrum: 
\begin{equation}
H_{\rm ra/er} = (I_{\circlearrowleft}^{\rm ra/er}-I_{\circlearrowright}^{\rm ra/er}),
\end{equation}
\noindent where "ra" and "er" denote the intraband and interband current contributions, respectively.~This quantity allows us to track both the dominant emission channel and its polarization imbalance as a function of the time delay between the pulses. The results for graphene and Bi$_2$Se$_3$ are shown in Fig.~\ref{Disc1}. In panel a), we present the helicity contributions for graphene for the third, fifth, and seventh harmonics, separating intraband (solid lines) and interband (dashed lines) contributions. Panel b) shows the corresponding results for Bi$_2$Se$_3$.~We observe that the intraband and interband contributions exhibit a similar dependence on the time delay, indicating that both channels contribute to the observed helicity modulation. 

\noindent For the lowest harmonic orders, such as the third harmonic, the intraband contribution is typically dominant, whereas for higher harmonics, the interband processes become increasingly important.~This crossover reflects the different physical origins of low- and high-energy emission channels in solids.~Furthermore, the inter and intraband helicity, $H_{\rm ra/er}$ aligns well with the description of the intraband contribution typically being dominant for the low harmonics, as shown for the third harmonic in Fig.~\ref{Disc1}(a).~Moreover, for the higher harmonics, the dominant contribution is the interband, as shown for the seventh harmonic in the same panel.~Additionally, an imprint of the underlying crystal symmetry is visible in the helicity dynamics, where successive harmonics exhibit alternating polarization behavior consistent with parity constraints (sign of $H_{ra/er}$).~Similar behaviour was observed for other materials (see Supplementary material).~The quantity $H_{\rm ra/er}$ reflects the continuous transition between linear-  (for $\delta t \approx T_0/2$ or $T_0$) and circular-like polarization states (for $\delta t \approx T_0/4$ or $3T_0/4$).~In the linear-like regime, both helicity components contribute equally, whereas in the circular-like regime one helicity channel becomes dominant depending on the sign of $\delta t$. The modulation of the driving field therefore induces a corresponding modulation of the interband and intraband currents, which manifests as a controllable helicity of the emitted harmonics. This behavior reflects the sensitivity of nonlinear emission to the sub-cycle structure of the light-matter interaction.

\noindent
A central aspect of high-order harmonic generation in solids is the role of dynamical symmetries, which combine spatial rotations with temporal translations, $\hat{U}=\hat{R}_{2\pi/n}\hat{T}_{T/n}$, and determine both the allowed harmonic orders and their polarization properties. For circularly polarized driving fields, this symmetry enforces strict selection rules, restricting emission to harmonics of order $q = nj \pm 1$ with alternating helicity. In this case, the driving field carries a well-defined angular momentum, and the light--matter interaction preserves the corresponding rotational symmetry.

\noindent
Polarization-crafted beams fundamentally modify this picture. Due to the finite time delay between orthogonally polarized components, the driving field can be expressed as a coherent superposition of co- and counter-rotating circular components with time-delay dependent amplitudes. Even when one helicity dominates, the residual counter-rotating contribution breaks the exact dynamical symmetry associated with ideal circular polarization. As a result, the symmetry operator $\hat{U}$ no longer leaves the Hamiltonian invariant, $\hat{U}\hat{H}(t)\hat{U}^\dagger \neq \hat{H}(t)$, and the corresponding selection rules are relaxed.

\noindent
This controlled symmetry breaking redistributes the emission among the allowed helicity channels without modifying the underlying crystal symmetry. Consequently, while the material determines the set of symmetry-allowed harmonic orders, the driving field governs the relative population and helicity content of these channels. This establishes harmonic helicity as a field-controlled observable arising from the tunable imbalance between co- and counter-rotating components of the driving field.

\emph{Summary--} We have demonstrated deterministic control of harmonic helicity in solids driven by polarization-crafted beams. By introducing a controlled time delay between orthogonally polarized fields, the polarization state of individual harmonics can be continuously tuned from linear to circular. This behavior is robust across materials with distinct symmetry and topology, indicating that the helicity of the emitted radiation is predominantly governed by the driving field. 

\noindent
We show that this control originates from a tunable imbalance between co- and counter-rotating circular components of the driving field, induced by the time delay. This imbalance leads to a controlled breaking of the dynamical symmetry associated with circular polarization and results in a redistribution of emission among helicity channels. By resolving interband and intraband contributions, we identify the microscopic origin of this effect as a sub-cycle modulation of the dipole coupling.

\noindent
These results establish harmonic helicity as a field-controlled observable that can be engineered independently of the material. Our work positions polarization-crafted beams as a versatile platform for generating tailored polarization states in HHG, enabling new opportunities for polarization-controlled attosecond light sources. Furthermore, since helicity is not a topological invariant quantity, our findings highlight the need for careful interpretation of helicity-resolved measurements as probes of topological matter.

\section{Acknowledgment}

\noindent C.B., J.M., and A.C. acknowledge the Sistema Nacional de Investigaci\'on de Panam\'a for partial financial support.~We thank~Prof.~M.~Santamar\'ia and Prof.~N.~Correa for logistical support at the Centro de Investigaci\'on con T\'ecnicas Nucleares, Universidad de Panam\'a. 
This work was supported by the Institute for Basic Science (IBS), Korea under Project Code IBS-R014-A1. D.K. acknowledges support from the National Research Foundation of Korea grants (Grant no. NRF-2023R1A2C2007998). This study was also supported by the MSIT (Ministry of Science and ICT), Korea, under the ITRC (Information Technology Research Center) support program (Grant no. IITP-2023-RS-2022-00164799) supervised by the IITP (Institute for Information Communications Technology Planning Evaluation).
W. Gao thanks financial support from National Natural Science Foundation of China (Grant number: 12474309).

\bibliographystyle{apsrev4-2}
\bibliography{references}


\section{Theoretical framework}

\noindent We calculate the HHG spectrum using the time-dependent density matrix (TDDM) formalism, following the approach presented in Refs.~\cite{KimSym2022,KimJourPhy2023}.~To simulate the electron dynamics in the solid driven by the strong, ultrashort external electromagnetic field, $\mathbf{E}(t)$, we employ the TDDM. The equations describe the time evolution of the reduced density matrix, $\hat{\rho}(\textbf{K},t)$, in the presence of the external field. Also $\textbf{P}_{mn}$ are the kinetic momentum matrix elements of the momentum operator $\hat{\textbf{P}}$ in the $\textbf{k}$--crystalline momentum space or Brillouin zone BZ.

\noindent Additionally, we use the moving frame to describe the macroscopic laser-matter interaction, in which the TDDMs take the following mathematical form \cite{ChaconPRB2020,YuePRA2020, VirkPRB2007}:
\begin{eqnarray}
    i \frac{\partial}{\partial t} \rho_{mn}^{(\mathrm{H})}(\mathbf{K}, t) &=& \left[ H_0^{(\mathrm{H})}(\mathbf{K} + \mathbf{A}(t)), \rho^{(\mathrm{H})}(\mathbf{K}, t) \right]_{mn} \nonumber \\
    &+& \mathbf{E}(t) \cdot \left[ \mathbf{D}^{(\mathrm{H})}(\mathbf{K} + \mathbf{A}(t)), \rho^{(\mathrm{H})}(\mathbf{K}, t) \right]_{mn} \nonumber \\
    &-& i (1 - \delta_{mn}) \frac{\rho_{mn}^{(\mathrm{H})}}{T_2}.\label{SBE}
\end{eqnarray}
\noindent In Eq.~(\ref{SBE}), the superscript $(\mathrm{H})$ denotes quantities 
expressed in the Hamiltonian (eigenstate) gauge.~The material-dependent terms are represented by the unperturbed Hamiltonian $H_0(\mathbf{k})$ and the dipole matrix $\mathbf{D}_{mn}^{(\mathrm{H})}(\mathbf{k}) = -i\langle u_m | \partial_{\mathbf{k}} | u_n \rangle$. Here, the subscripts $m$ and $n$ describe the valence and conduction bands, respectively. The Berry connection is described when $n=m$ and transition dipole matrix elements for $n\neq m$. The vector potential and electromagnetic field are represented by $\mathbf{A}(t)$ and $\mathbf{E}(t)$, respectively. The final term describes the decay of the system back to its ground state, where $T_2$ is the dephasing time. This term usually condenses high order electron correlations and is taken to be around half of the laser period. 

\noindent The numerical solution of Eq.(\ref{SBE}) requires continuous quantities such as transition dipole matrix elements and Berry connections. However, in topological materials (as well as some trivial ones), a globally smooth wavefunction (amplitude and phase) gauge may not exist \cite{KOHMOTO1985343}.~This introduces numerical artifacts, including artificial singularities in the currents \cite{ChaconPRB2020}.~To circumvent this problem, we employ the Wannier gauge, constructing a Bloch-like basis from localized Wannier functions \cite{SilvaPRB2019}. In this representation, the tight-binding Hamiltonian $H_0(\mathbf{k})$ and the dipole matrix $\mathbf{D}^{(\mathrm{W})}_{mn}(\textbf{k})$ remains smooth (free of singular points). Here, the superscript $(\mathrm{W})$ denotes quantities expressed in the Wannier gauge.

\noindent Explicitly, the Wannier functions are defined as $w_m(\mathbf{r} - \mathbf{R}) = \langle \mathbf{r} | \mathbf{R} m \rangle$, where $\mathbf{R}$ is the Bravais lattice vector. Assuming an orthogonal basis $\langle \mathbf{R}_m | \mathbf{R}'_n \rangle = \delta_{m,n} \delta_{\mathbf{R}, \mathbf{R}'}$, a Bloch-like basis can be constructed as follows:
\begin{align}
|\psi^{(\mathrm{W})}_{n{\mathbf{k}}}\rangle = \frac{1}{\sqrt{N_c}} \sum_{\mathbf{R}} 
e^{i \mathbf{k} \cdot (\mathbf{R} + \Delta_n)} |\mathbf{R} n \rangle,
\end{align}
\noindent 
\noindent where $N_c$ is the number of unit cells and $\Delta_n = \langle \mathbf{0}_n | \hat{\mathbf{r}} | \mathbf{0}_n \rangle$ represents the Wannier center, which in our model corresponds to the position of the atom in the unit cell. Furthermore, assuming that the Wannier functions are well-localized, we approximate: $\langle \mathbf{R}_m | \hat{\mathbf{r}} | \mathbf{0}_n \rangle = \Delta_n \delta_{mn}$, which eliminates the transition matrix elements and Berry curvature terms, so that $\mathbf{D}_{mn}^{(\mathrm{W})}(\mathbf{k})={\textbf{0}}$ in Eq.~(\ref{SBE}) \cite{KimSym2022}.~Consequently, the TDDMs in the Wannier basis reduce to:
\begin{eqnarray}
         i \frac{\partial}{\partial t} \rho^{(\mathrm{W})}_{nm}(\mathbf{K}, t) &=& \left[ H_0^{(\mathrm{W})}(\mathbf{K} + \mathbf{A}(t)), \rho^{(\mathrm{W})}(\mathbf{K}, t) \right]_{nm}. \nonumber \\
    \label{SBEW}
\end{eqnarray}
\noindent Since the dephasing term $T_2$ is not defined in the Wannier gauge, the density matrix must be transformed to the Hamiltonian gauge via a unitary transformation. Within this framework, the dephasing term is treated in the eigenstate basis, while the remaining terms are computed in the Wannier basis. The relationship between both gauges is given by:
\begin{equation}
    \rho^{(\mathrm{H})}_{nm}(\mathbf{K}, t)= \sum_{a,b} U^{\dagger}_{nb}(\mathbf{K})\rho^{(\mathrm{W})}_{ba}(\mathbf{K}, t)U_{am}(\mathbf{K})
    \label{eq:rho_transformation_H},
\end{equation}
\begin{equation}
    \rho^{(\mathrm{W})}_{nm}(\mathbf{K}, t)= \sum_{ab} U_{nb}(\mathbf{K})\rho^{(\mathrm{H})}_{ba}(\mathbf{K}, t)U^{\dagger}_{am}(\mathbf{K}),
\end{equation}
\noindent where $a,b$ indices refer to the Wannier site-orbital basis and $n,m$ indices refer to the Hamiltonian eigenstate basis. Importantly, for topological materials, the multiband formulation is chosen so that the net topological charge cancels identically obtaining smooth, well defined maximally localized Wannier states \cite{vanderbilt2018berry} and the calculation of topological quantities is done in the eigenstate basis. 

\noindent For the harmonic generation process, we solve Eq.~(\ref{SBEW}) to find the particular density matrix of the system, $\hat{\rho}(\textbf{K},t)$, followed by calculating the generated current in the solid-material, ${\textbf{J}}(t)$. The current can be calculated from the following expression:
\begin{equation}
    \mathbf{J}(t) = \sum_{mn} \int_{\textrm{BZ}} d\textbf{K} \, \mathbf{P}_{mn}^{(\textrm{W})}(\mathbf{K} + \mathbf{A}(t)) \, \rho_{nm}^{(\textrm{W})}(\mathbf{K}, t),
    \label{eq:Jtotal}
\end{equation}
\noindent where ${\bf P}_{mn}^{(\textrm{W})}({\bf k})$ is the momentum matrix element in the Wannier gauge, which takes the following form: 
\begin{equation}
    {\bf P}_{mn}^{(\mathrm{W})}({\bf k}) = \langle \psi_{m{\bf k}}^{(\mathrm{W})} | 
    \partial_{\bf k}\hat{H}_0^{(\mathrm{W})}({\bf k}) | \psi_{n{\bf k}}^{(\mathrm{W})} \rangle .
\end{equation}
\noindent For a better understanding of the physical mechanisms behind the HHG dynamics, the total current ($\ref{eq:Jtotal}$) can be divided into two terms: the interband current and the intraband current. Since these are originally defined in the Hamiltonian gauge using the eigenstates as a basis, we apply the unitary transformation in Eq.~(\ref{eq:rho_transformation_H}) so that the contributions can be expressed as:
\begin{equation}
    \mathbf{J}_ \text{er}(t) = \sum_{n \neq m} \int_{\text{BZ}} \text{d}\mathbf{K} \mathbf{P}_{nm}^{(\text{H})}(\mathbf{K} + \mathbf{A}(t)) \rho_{nm}^{(\text{H})}(\mathbf{K}, t),
\end{equation}
\begin{equation}
    \mathbf{J}_ \text{ra}(t) = \sum_{m} \int_{\text{BZ}} \text{d}\mathbf{K} \mathbf{P}_{mm}^{(\text{H})}(\mathbf{K} + \mathbf{A}(t)) \rho_{mm}^{(\text{H})}(\mathbf{K}, t).
\end{equation}
\noindent The total current is thus given by the sum of both contributions:
\begin{equation}
    \mathbf{J}(t) = \mathbf{J}_{\text{er}}(t) + \mathbf{J}_{\text{ra}}(t) \label{TC}.
\end{equation}
\noindent Finally, the HHG spectrum is computed by Fourier transforming the time derivative of the current given in Eq.~(\ref{TC}):
\begin{equation}
    I_{\textrm{HHG}}(\omega) = \left| \mathcal{FT} \left[ \frac{d}{dt} \mathbf{J}(t) \right] \right|^2. \label{TYield}
\end{equation}

\section{Dipole coupling}

\noindent To further understand the differences in the harmonic spectra from the RCP/LCP and PCBs, we analyze two key aspects of the light-matter interaction: the Bloch acceleration theorem (BAT) and the coupling of the laser field to the dipole matrix elements. The BAT describes the evolution of an electron wave packet in a periodic potential as $\hbar \dot{\mathbf{k}}_c(t) = -e\mathbf{E}(t)$. For circularly polarized driving fields, this leads to trajectories of the form:

\begin{equation}
    \hbar\vec{k}_c(t) =k_{0x}\hat{x} -\frac{eE_0}{\omega_0}\sin(\omega_0 t)\hat{x} + k_{0y}\hat{y} \pm \frac{eE_0}{\omega_0}\cos(\omega_0 t)\hat{y}, \label{BATC}
\end{equation}
where, for simplicity we assumed an  envelope phase equal to zero.~For PCBs, however, the time delay provides a means to control the electron dynamics, as follows:
\begin{eqnarray}
    \hbar\vec{k}_c^{\rm (PCB)}(t)  &=&\vec{k}_{0x} -\frac{eE_0}{\omega_0}\sin(\omega_0 t)\hat{x} + \vec{k}_{0y} \nonumber \\
   & -& \frac{eE_0}{\omega_0}\sin(\omega_0 (t-\delta t))\hat{y}. \label{BATPBC}
\end{eqnarray}
\noindent In this case, the time delay $\delta t$ controls the polarization state of the field, continuously spanning from linear to circular polarization.~However, since the resulting trajectories are related by a phase shift in the field components, the BAT alone cannot explain the observed helicity-dependent harmonic response in the materials considered here.

\noindent The key mechanism instead arises from the light-matter coupling via the dipole matrix elements, $\mathbf{E}(t)\cdot \mathbf{d}_{cv}$. For circularly polarized fields, this coupling selects well-defined helicity channels,
\begin{eqnarray}
    \text{RCP} &&\rightarrow \frac{E_0}{2}e^{i\omega_0t}\hat{e}_+\cdot\vec{d}_{cv}^-(\mathbf{k}) \nonumber \\
    \text{LCP} &&\rightarrow \frac{E_0}{2}e^{i\omega_0t}\hat{e}_-\cdot\vec{d}_{cv}^+(\mathbf{k}), \label{DMcoupling}
\end{eqnarray}
\noindent leading to harmonics with well-defined polarization. Note that we have written the dipole moment matrix as $d^{\pm}_{cv}(\mathbf{k})=d_{cv}^{(\rm x)}(\mathbf{k})\hat{x} \pm id_{cv}^{(\rm y)}(\mathbf{k})\hat{y}$.~For PCBs, the field can be decomposed into circular components of the form $E_{\pm} \propto (1 \pm ie^{i\omega_0 \delta t})$, which leads to the coupling:
\begin{equation}
\mathbf{E}(t)\cdot \mathbf{d}_{cv}(\mathbf{k}) \propto (1 - i e^{i\omega_0\delta t}) d_-(\mathbf{k}) + (1 +i e^{i\omega_0 \delta t}) d_+(\mathbf{k}).
\end{equation}
This expression shows that the relative weight of the two helicity channels is directly controlled by the time-delay $\delta t$. In particular, $\delta t = T_0/4$ leads to nearly pure helicity, while $\delta t = 0$ also yields linear polarization.
Equivalently, in the linear basis, the coupling reads
\begin{eqnarray}
\mathbf{E}(t)\cdot \mathbf{d}_{cv}(\mathbf{k}) &=& E_0\cos(\omega_0 t) d_{x}+E_0d_y\Big[ \cos(\omega_0 t)\cos(\omega_0\delta t) \nonumber \\
&+& \sin(\omega_0 t)\sin(\omega_0 \delta t)\Big],
\end{eqnarray}
\noindent which shows explicitly how the delay modulates the relative field components. The instantaneous coupling strength differs for opposite helicities. For example, at $\delta t = \pm T_0/4$:
\begin{eqnarray}
|\mathbf{E}(t)\cdot \mathbf{d}_{cv}(\mathbf{k})|^2 &\propto& E_0^2 (\cos^2 (\omega_0 t) |d_x|^2 + \sin^2(\omega_0 t)|d_y|^2 \nonumber \\
&\pm& 2\cos(\omega_0 t)\sin(\omega_0 t)\mathrm{Re}(d_xd_y^*),
\end{eqnarray}
which reveals that helicity dependence arises from the interference between orthogonal dipole components, governed by $\mathrm{Re}(d_x d_y^*)$. The sign change in this cross term for opposite delays leads to a sub-cycle modulation of the coupling, while the cycle-averaged intensity remains unchanged. This modulation directly tracks the sub-cycle evolution of the field helicity, as shown in Fig.~\ref{TDHelicity}.

These results indicate that helicity control originates from the time-delay dependent modulation of the dipole coupling induced by the polarization-crafted fields, with additional phase-dependent corrections from the Bloch acceleration dynamics. Since the effect relies on sub-cycle dynamics, it diminishes for longer pulses where temporal asymmetries are averaged out.

The interband dipole matrix elements~$\mathbf{d}_{cv}(\mathbf{k}) = -i\langle u_c(\mathbf{k})|\nabla_{\mathbf{k}}|u_v(\mathbf{k})\rangle$ inherit the geometric phase structure of the Bloch states, making the interference term sensitive to band geometry, including Berry connection effects. As a result, helicity control constitutes a broadly applicable, field-driven mechanism, while its magnitude and spectral response remain material-dependent through the amplitude and phase structure of the dipole matrix elements and the band structure.

\section{Helicity decomposition of polarization-crafted beams}

\noindent
To further clarify the origin of helicity control, we derive an explicit decomposition of the polarization-crafted beam (PCB) into circular polarization components directly in the time domain. Starting from Eq.~(\ref{MathPCBs}), and assuming equal amplitudes and carrier-envelope phases for simplicity, the field can be written as
\begin{equation}
\mathbf{E}(t) = f(t)\cos(\omega_0 t)\,\hat{x} + f(t-\delta t)\cos[\omega_0(t-\delta t)]\,\hat{y}.
\label{Epcb_start}
\end{equation}

\noindent
For delays $\delta t = \pm T_0/4$, with $T_0 = 2\pi/\omega_0$, we use
\begin{equation}
\cos[\omega_0(t-\delta t)] = \cos\left(\omega_0 t \mp \frac{\pi}{2}\right) = \pm \sin(\omega_0 t),
\end{equation}
to obtain
\begin{equation}
\mathbf{E}(t) = f(t)\cos(\omega_0 t)\,\hat{x} \pm f(t\mp T_0/4)\sin(\omega_0 t)\,\hat{y}.
\label{Epcb_shift}
\end{equation}

\noindent
We now introduce the circular polarization basis vectors
\begin{equation}
\mathbf{e}_{\pm}(t) = \cos(\omega_0 t)\hat{x} \pm \sin(\omega_0 t)\hat{y},
\end{equation}
which correspond to right- and left-rotating fields. Using these definitions, Eq.~(\ref{Epcb_shift}) can be rewritten exactly as
\begin{align}
\mathbf{E}(t) &= \frac{f(t)+f(t\mp T_0/4)}{2}\,\mathbf{e}_{\pm}(t) \nonumber \\
&+ \frac{f(t)-f(t\mp T_0/4)}{2}\,\mathbf{e}_{\mp}(t).
\label{Epcb_decomp}
\end{align}

\noindent
This expression shows that a PCB with delay $\delta t = \pm T_0/4$ is a coherent superposition of co- and counter-rotating circular components with amplitudes
\begin{eqnarray}
A_{co}(t) = \frac{f(t)+ f(t- T_0/4)}{2}\nonumber \\
A_{counter}(t) = \frac{f(t)- f(t- T_0/4)}{2}.
\end{eqnarray}

\noindent
Under the slowly varying envelope approximation, $f(t\mp T_0/4) \approx f(t)$, such that $|A_{co}(t)| \ll |A_{counter}(t)|$, and the field is dominated by a single helicity component. However, the counter-rotating contribution remains finite and introduces a controlled deviation from perfect circular polarization. Importantly, this residual component breaks the exact dynamical symmetry associated with circularly polarized driving fields. As a result, the light--matter Hamiltonian is no longer invariant under the combined symmetry operation $\hat{U}=\hat{R}_{2\pi/n}\hat{T}_{T/n}$, which in turn relaxes the strict helicity selection rules. The emitted harmonic helicity is therefore determined by the imbalance between the two components,
\begin{equation}
\Delta H(t) \propto |A_{co}(t)|^2 - |A_{counter}(t)|^2,
\end{equation}
providing a direct link between the time delay $\delta t$ and the helicity of the emitted radiation.

\noindent
This time-domain picture is fully consistent with the frequency-domain description of the dipole coupling, where the PCB field introduces relative weights $(1 \pm i e^{i\omega_0\delta t})$ between the two helicity channels. The delay $\delta t$ therefore acts as a control parameter that continuously tunes the relative population of co- and counter-rotating emission pathways.

\end{document}